\documentclass{aastex631}

\usepackage{amsmath}
\usepackage{amssymb}
\usepackage{graphicx}
\usepackage{CJKutf8}
\usepackage{epstopdf}
\usepackage{hyperref}
\usepackage{array}
\usepackage{rotating}

\begin{document}
\begin{CJK}{UTF8}{gbsn}

\title{Local Particle Acceleration in an ICME-in-Sheath Structure Observed by Solar Orbiter}
\correspondingauthor{Chuan Li}
\email{lic@nju.edu.cn}

\author[0009-0003-3760-705X]{Xiaomin Chen}
\affiliation{School of Astronomy and Space Science, Nanjing University, Nanjing 210023, People’s Republic of China}
\affiliation{Key Laboratory of Modern Astronomy and Astrophysics (Nanjing University), Ministry of Education, Nanjing 210023, People’s Republic of China} 

\author[0000-0001-7693-4908]{Chuan Li}
\affiliation{School of Astronomy and Space Science, Nanjing University, Nanjing 210023, People’s Republic of China}
\affiliation{Key Laboratory of Modern Astronomy and Astrophysics (Nanjing University), Ministry of Education, Nanjing 210023, People’s Republic of China} 
\affiliation{Institute of Science and Technology for Deep Space Exploration, Suzhou Campus, Nanjing University, Suzhou 215163, People’s Republic of China} 

\author[0000-0002-9246-996X]{Zigong Xu}
\affiliation{California Institute of Technology, MC 290-17, Pasadena, CA 91125, USA} 

\author[0000-0003-3623-4928]{Georgios Nicolaou}
\affiliation{Department of Space and Climate Physics, Mullard Space Science Laboratory, University College London, Holmbury St Mary, Dorking, Surrey RH5 6NT, UK}

\author[0000-0002-9471-5132]{Alexander Kollhoff}
\affiliation{Institute of Experimental and Applied Physics, Kiel University, Germany}

\author[0000-0003-1093-2066]{George C. Ho}
\affiliation{Southwest Research Institute, San Antonio, TX 78228, USA}

\author[0000-0002-7388-173X]{Robert F. Wimmer-Schweingruber}
\affiliation{Institute of Experimental and Applied Physics, Kiel University, Germany}

\author[0000-0002-5982-4667]{Christopher J. Owen}
\affiliation{Department of Space and Climate Physics, Mullard Space Science Laboratory, University College London, Holmbury St Mary, Dorking, Surrey RH5 6NT, UK}

\begin{abstract}

Local particle acceleration in the shock sheath region formed during the interaction between multiple coronal mass ejections (CMEs) is a complicated process that is still under investigation. On March 23, 2024, the successive eruption of two magnetic flux ropes (MFRs) from the solar active region 3614 produced twin CMEs, as identified in coronagraph images. By analyzing in-situ data from Solar Orbiter and Wind, it is found that the primary ICME-driven shock overtook the preceding ICME, trapping it in the sheath between the shock and the primary ICME, forming the ICME-in-sheath (IIS) structure. Using Solar Orbiter observations, we show that both electrons and ions are accelerated within the IIS. A clear enhancement of suprathermal electrons was observed at the IIS boundary, where strong flow shear and large magnetic field variation suggest possible local electron acceleration. Electrons ($\textgreater$38 keV) exhibit a long-lasting enhancement in the IIS with a spectral index of $\sim$2.2, similar to that in the shock sheath and the primary ICME, indicating a similar solar origin. Inside both the sheath and IIS, spectra of proton and $^4$He are generally consistent with the prediction of the diffusive shock acceleration, whereas Fe and O present a double power-law shape. Additionally, the Fe/O ratio in the IIS is higher than that in the sheath, and more close to the abundance of the flare-related particles, suggesting the remnant particles of flare confined in the IIS.           
\end{abstract}


\section{Introduction} \label{sec:intro}

Fast forward shocks driven by interplanetary coronal mass ejections (ICMEs) deflect and compress the ambient solar wind during its propagation, forming a sheath region between the shock and the magnetic cloud. This turbulent sheath region is characterized by elevated density and temperature \citep{sheath_review}. As the shock propagates from the corona to the interplanetary space, a portion of the accelerated particles are trapped in the shock downstream, resulting in an enhancement of the energetic particle flux in the sheath region. This enhancement is known as energetic storm particles (ESPs)\citep{esp1,Reames}. The sheath structure plays a crucial role in shaping the particle behavior of ESPs. Observations by \cite{esp_mfr} have revealed small-scale flux ropes and embedded current sheets within the sheath, which are associated with the enhancement of energetic ions. At the boundary between the sheath and the ICME, large magnetic shear between the closed magnetic field of the ICME and the open field of the sheath may lead to magnetic reconnection \citep{bl_reconnection}. Furthermore, Wind and ACE have observed local electron acceleration at this boundary, driven by electric field acceleration and Fermi-type processes during magnetic reconnection \citep{prl}.

The CME-driven shocks may interact with magnetic structures in the heliosphere, such as the high-speed streams \citep{interaction_hss}, co-rotating interaction regions \citep{interaction_cir} and preceding CMEs \citep{Lugaz}. The complex interaction processes may influence the sheath structure and the associated ESPs considerably. In a shock event observed by Parker Solar Probe (PSP), \cite{Giacalone1} interpreted the ion dropout of ESPs as the interaction between the shock and an isolated magnetic structure. In particular, the interaction between successive CMEs is an increasingly important issue for space weather forecast. Recently, \cite{interaction_chi} reported the direct observation of a CME-driven shock propagating through two preceding CMEs, which resulted in the compression of the preceding ICME and the decrease of the shock strength. \cite{iis} proposed the ICME-in-Sheath (IIS) structure, which described the scenario in which a primary ICME-driven shock crosses a preceding ICME, causing the preceding ICME to be stuck in the sheath between the shock and the primary ICME. After being compressed and modified by the primary shock, the IIS often loses its typical ICME signatures. Compared to typical ICMEs, the IIS exhibits enhanced magnetic field, density and temperature. The magnetic field within the IIS may also display turbulent characteristics, similar to that in the shock sheath.    

The interaction of CMEs plays an important role in large solar energetic particle (SEP) events. The type-II radio enhancement associated with interacting CMEs, as proposed by \cite{type_ii_g}, provides evidence that CME interactions can strengthen shocks and thereby enhance SEP acceleration. Several SEP events associated with multiple CMEs were analyzed by comparing the particle temporal profile with the coronagraph images and radio signals \citep{sep_inteaction_li,interaction_chen}. Meanwhile, the statistic study of \cite{interaction_g} showed that a fast CME with a preceding CME is more likely to be SEP-rich. To account for SEP observations during CME–CME interactions, \cite{twin_CME} proposed the twin-CME scenario to explain how the CME-CME interaction near the Sun could enhance both the seed population and turbulent level for the primary CME-driven shock acceleration. \cite{interaction_Niemela} performed a simulation of particle acceleration associated with merging CME-driven shocks and suggested that the ``macroscopic collapsing magnetic trap" effect leads to more efficient first-order Fermi acceleration. Observations of energetic particle enhancement in the shock-ICME complex structures (S-ICMEs) further indicate that particles can be accelerated during shock-ICME interactions and subsequently confined in the S-ICMEs \citep{s_icme_shen,s_icme_xu}. Nevertheless, more efforts are still needed to understand the detailed acceleration processes during CME-CME interaction. 

In this study, we investigate an IIS structure and associated particle acceleration processes during a twin-CME event on 2024 March 23 observed by Solar Orbiter and Wind. Section \ref{sec:instrument} introduces the instruments and datasets used in this study. Section \ref{sec:observation} presents remote-sensing observation of the solar eruptions and the in-situ measurements of the ICMEs, followed by an analysis of electron and ion acceleration within the sheath and the IIS. Section \ref{sec:summary} provides a brief conclusion and discussion.

\section{Instrumentation} \label{sec:instrument}

In this study, we use the in-situ data from Solar Orbiter and Wind. The Magnetometer (MAG, \citealt{mag}) onboard Solar Orbiter provides the magnetic field measurement. The Solar Wind Analyser (SWA, \citealt{swa}) consists of three detectors: an Electron Analyser System (EAS), a Proton-Alpha Sensor (PAS) and a Heavy-Ion Sensor (HIS). We use the proton plasma parameters derived from the Level 2 ground-calibrated moments of the proton vdfs observed from SWA-PAS, and derive the pitch angle distributions (PADs) of electrons for this study by using SWA-EAS and MAG measurements. The phase space density of electrons from EAS in the energy range of $\sim$1 eV to 5 keV is analyzed in this study. The energetic particle measurement is obtained from the Energetic Particle Detector (EPD, \citealt{epd}), which consists of four instruments: Suprathermal Electrons and Protons (STEP), Electron Proton Telescope (EPT), High Energy Telescope (HET), and Suprathermal Ion Spectrograph (SIS). We use the Level 3 EPT electron data, in which the ion contamination on the electron has been corrected(\href{https://data.serpentine-h2020.eu/l3data/solo/}{https://data.serpentine-h2020.eu/l3data/solo/}). We analyze the differential energy flux of electrons observed by EPT over an energy range of $\sim$30 keV to 200 keV. The measurements of H, $^4$He, Fe and O are provided by SIS, covering energies from a few hundred keV/n to several MeV/n. For in-situ measurement at L1 point, we use the magnetic field and solar wind data from the Wind spacecraft. The pitch angle distribution of electrons at L1 point is obtained from the electron energy-angle distribution (ELPD) data of 3D Plasma and Energetic Particle (3DP, \citealt{wind}) instrument onboard Wind.

To investigate the solar eruption, we analyze the extreme ultraviolet (EUV) images from the Atmospheric Imaging Assembly (AIA, \citealt{aia}) onboard Solar Dynamics Observatory (SDO, \citealt{sdo}) and the H$\alpha$ images from the H$\alpha$ Imaging Spectragraph (HIS, \citealt{his}) of the Chinese H$\alpha$ Solar Explorer (CHASE, \citealt{chase}). The white-light observation of the CME is obtained from the Large Angle and Spectrometric Coronagraph (LASCO, \citealt{lasco}) onboard the Solar and Heliospheric Observatory (SOHO). 
    
\section{Observations and Results} \label{sec:observation}
\subsection{Overview of the solar eruption}

\begin{figure}[htbp]
\centering
\includegraphics[width=1\linewidth]{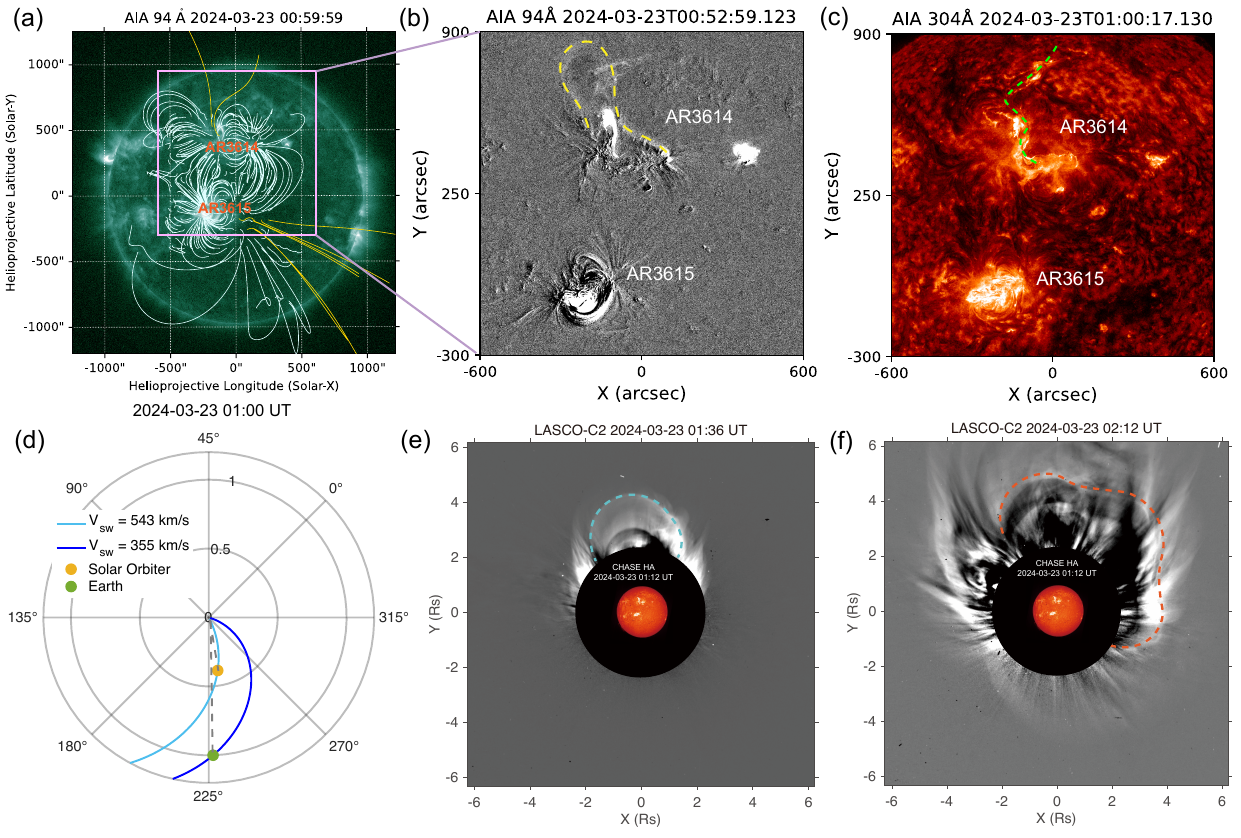}
\caption{\label{fig:overview} Overview of the solar eruption on 2024 March 23. (a) PFSS modeling of magnetic fields overlaid on the AIA 94\AA~image, with closed and open field lines shown in white and yellow, respectively. (b) Base-difference AIA 94\AA~image, with the magnetic flux rope outlined by a yellow dashed line. (c) AIA 304\AA~image, highlighting the erupting filament in green. (d) Positions of Solar Orbiter and Earth on the ecliptic plane with IMF Parker spirals. The solar wind speed is averaged over 00:00–01:00 UT and indicated in the legend. (e) LASCO-C2 white-light and CHASE H$\alpha$ images. The CHASE H$\alpha$ image was taken at 01:12 UT during the impulsive phase of the flare. (f) Same as (e), but for the second CME.}
\end{figure}

Fig. \ref{fig:overview} presents an overview of the solar eruption on 2024 March 23. Panel (a) shows the potential field source surface (PFSS) modeling of coronal magnetic fields overlaid on the AIA 94\AA~image, which was obtained using the pfsspy package \citep{pfsspy}. The X1.1 class flare started at 00:58 UT, peaked at 01:33 UT and stopped at 01:52 UT. The source region is two magnetically connected active regions, AR3614 and AR3615, forming a quadrupolar magnetic configuration. EUV observations reveal that two successive eruptions of magnetic flux ropes (MFRs) occurred in AR3614. Among two MFRs, one exhibited a hot-channel structure, as shown in Panel (b), while the other one appeared as a large-scale filament, as depicted in Panel (c). Erupting nearly simultaneously at about 01:00 UT, the two MFRs produced two distinct CME components observed in coronagraph images, constituting a typical twin-CME scenario.    

The northward halo CME observed by LASCO-C2 at 01:36 UT is shown in Panel (e) of Fig. \ref{fig:overview}, has its bright front marked by a cyan dashed line. The second halo CME observed at 02:12 UT is marked by an orange line in Panel (f). A strong Forbush decrease was detected when the CMEs arrived at the Earth, which is reported by \cite{FB_decrease}. The authors also analyzed the EUV and white-light images and suggested a twin-CME scenario. In the LASCO CME catalog (\href{https://cdaw.gsfc.nasa.gov/CME_list/}{https://cdaw.gsfc.nasa.gov/CME\_list}), the twin CMEs are considered as a halo CME with a linear speed of 1470 km/s. Panel (d) illustrates the positions of Solar Orbiter and the Earth, where Solar Orbiter was at 0.4 AU and nearly radially aligned with the Earth. This configuration provides a valuable opportunity to investigate the interaction between twin CMEs and its influence on particle acceleration during their propagation from the corona to the Earth. 

\subsection{Structure of the shock sheath from 0.4 AU to 1 AU}

\begin{figure}[htbp]
\centering
\includegraphics[width=1\linewidth]{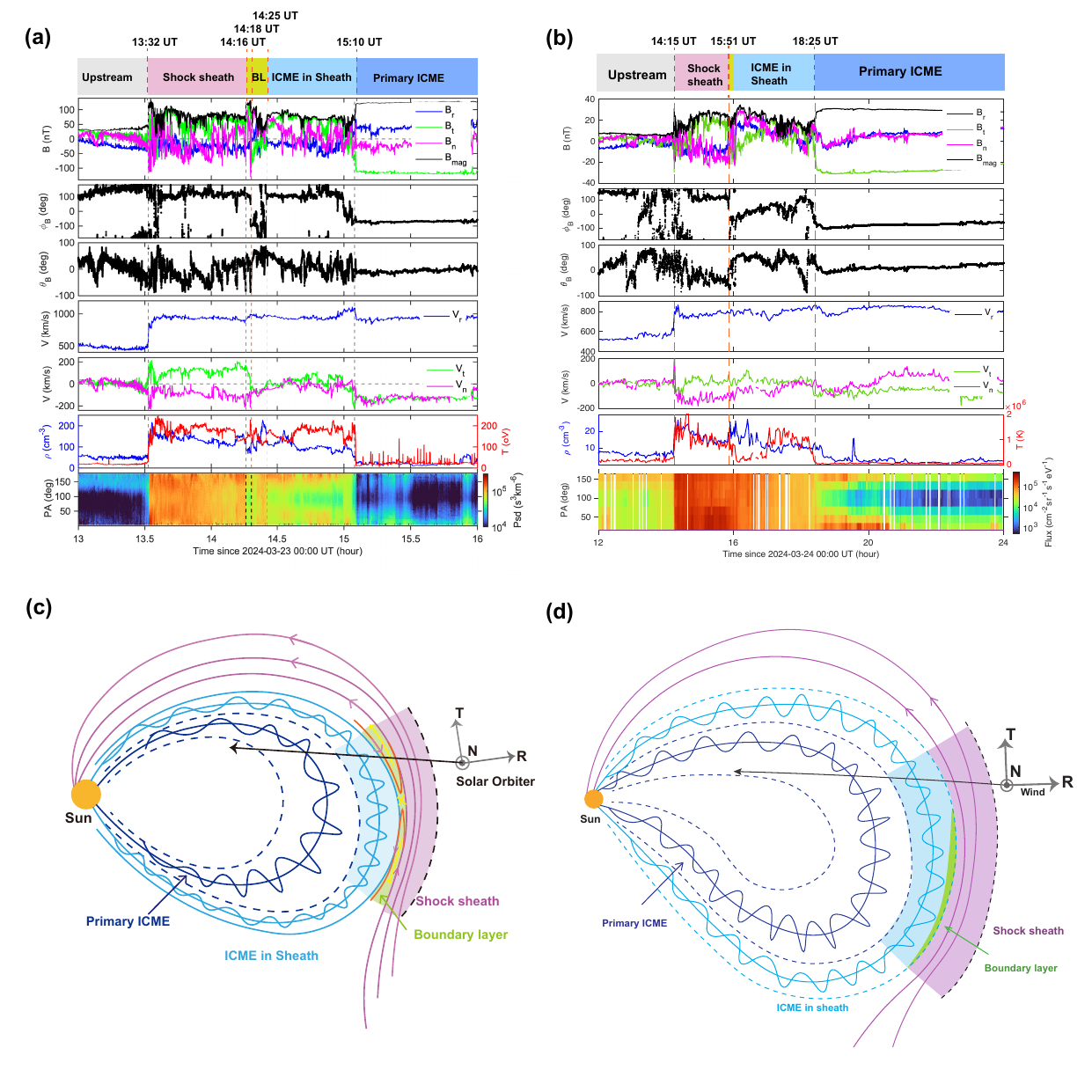}
\caption{\label{fig:sw_mag} In-situ measurements of the ICME and shock sheath structures. (a) Solar Orbiter measurements, showing, from top to bottom, the magnetic field in RTN coordinates, $\phi_B$, $\theta_B$, solar wind velocity, density, temperature, and PADs of strahl electrons ($\textgreater$70 eV). The five vertical dashed lines mark the shock arrival, the boundary layer of the IIS, a sharp rotation of magnetic field, the entering of the IIS, and the primary ICME. (b) Shock sheath parameters from Wind in RTN coordinates, with the last panel showing PADs of strahl electrons (121 eV). Three vertical dashed lines indicate the shock arrival, the IIS boundary, and the boundary between the IIS and the primary ICME. Panels (c) and (d) illustrate the crossings of Solar Orbiter and Wind of the shock sheath, the IIS and the ICME structures in the ecliptic plane.} 
\end{figure}

Panel (a) in Fig. \ref{fig:sw_mag} shows the in-situ measurement at Solar Orbiter at $\sim$0.4 AU. The CME-driven shock reached Solar Orbiter at 13:32 UT, $\sim$12 hours after the solar eruption, characterized by a sharp increase in magnetic field strength and solar wind velocity, density and temperature. The basic shock parameters, summarized in Table \ref{tab:shock}, were derived using the methods described in the Appendix \ref{sec:shock_parameters}, indicating a strong oblique shock with high compression ratio and Mach number. A normal part of the shock sheath, which will be called the shock sheath going forward, was detected from 13:32 UT to 14:16 UT. At 14:25 UT, the magnetic field magnitude became more stable compared to the shock sheath, while the $B_n$ component showed a smooth rotation. The presence of bi-directional streaming strahl electrons from 14:25 UT to 15:10 UT, a typical criterion for ICME identification, suggests that Solar Orbiter entered the main body of the IIS. During this interval, the magnetic field maintained a stable orientation along the tangential (T) direction, characterized by an azimuthal angle of around 90$^\circ$, which closely resembles the polarity observed in the shock sheath. This similarity could be related to the shock compressing and modifying the MFR structure of the IIS. Between the shock sheath and the IIS, a boundary layer was observed from 14:16 UT to 14:25 UT. It showed a reversed polarity and strong fluctuations in magnetic field strength. The $B_t$ component underwent a rapid polarity reversal, followed by a gradual decrease toward zero. Together with the large variations in solar wind properties  shown from the fourth to the bottom panels in Fig. \ref{fig:sw_mag}(a), these signatures indicate that Solar Orbiter was passing through multiple fine structures, which will be analyzed in detail below. At 15:10 UT, Solar Orbiter entered the primary ICME, characterized by an increase in total magnetic field magnitude and a rapid decrease in ion density and temperature. The duration of the primary ICME, through which Solar Orbiter passed, was $\sim$5 hours. The schematic in Panel (c) presents a possible scenario for the ICME and shock sheath structures, along with the trajectory of Solar Orbiter, which could explain the main observational features. The absence of clear evidence for a second shock within the sheath suggests that the shock driven by the primary CME probably had merged with the preceding CME-driven shock.


We then analyze the boundary layer of the IIS in detail. At 14:16 UT, marked by the first vertical orange line, the magnetic field magnitude increased sharply while the ion density decreased, resembling the characteristics of a magnetic cloud. At 14:18 UT, marked by the second vertical orange line, the total magnitude decreased, the $B_r$ component was $\sim$0, and the $B_t$ component reversed from 110 nT to -100 nT, while the $B_n$ component abruptly varied from $\sim$80 nT to $\sim$-140 nT and then returned to $\sim$80 nT, indicating a sharp magnetic field variation. A pronounced flow shear was observed, with $V_t$ changing from $\sim$140 km/s to $\sim$-50 km/s, suggesting the presence of a stream interface separating the sheath and the IIS. Simultaneously, both ion density and temperature increased, probably corresponding to a reconnection jet. This suggests that magnetic reconnection probably occurred at the interface between the sheath and the IIS, and Solar Orbiter likely crossed this region. Following this, as $B_t$ gradually decreased, Solar Orbiter entered a turbulent region characterized by low magnetic field magnitude and temperature but elevated ion density. 

At 14:15 UT on 2024 March 24, the shock arrived at Wind, as shown in Panel (b) of Fig. \ref{fig:sw_mag}. Using the same method described in Appendix \ref{sec:shock_parameters}, we derived shock parameters in Table \ref{tab:shock}. The results are generally consistent with the parameters in the CfA Interplanetary Shock Database (\href{https://lweb.cfa.harvard.edu/shocks/}{https://lweb.cfa.harvard.edu/shocks}). Compared to the shock observed by Solar Orbiter, the shock at 1 AU has larger Alfv$\acute{e}$n Mach number and similar density compression ratio, suggesting that the upstream Alfv$\acute{e}$n velocity decreased faster than the shock speed. At 15:51 UT, the magnetic field in RTN coordinates exhibited a rapid change in all directions, and a slight flow shear was observed. Between 15:51 UT and 18:25 UT, the magnetic field presented a smooth rotation, which fitted the characteristic of a MFR and was considered as the IIS. Within this interval, the azimuthal angle rotated from –50$^\circ$ to 80$^\circ$, while the elevation angle remained around 50$^\circ$. Compared with the IIS observed by Solar Orbiter, the structure detected by Wind showed a more pronounced twist, with the dominant magnetic component shifting from $B_t$ at 0.4 AU to $B_n$ at 1 AU, suggesting a possible rotation of the flux rope during its propagation. The proton density showed no obvious change when entering the IIS, while the proton temperature decreased significantly. At 18:25 UT, the spacecraft entered the primary ICME, marked by an increase in the total magnetic field and a decrease in both solar wind density and temperature. Within the primary ICME, strahl electrons exhibited clear bi-directional streaming. The duration of the primary ICME was $\sim$15.3 hours. The extended duration of the primary ICME and the little variation of $\phi_B$ and $\theta_B$ suggested that Wind probably crossed the leg of the primary ICME.


\setlength{\tabcolsep}{10pt} 
\renewcommand{\arraystretch}{1.5} 
\begin{table}[h!]
\centering
\begin{tabular}{ccccccc}
\hline
\textbf{} & \textbf{$n_{shock}$} & \textbf{$\theta_{Bn}$} & \textbf{$V_{shock}$} & \textbf{$V_{A_{up}}$} & \textbf{$M_A$} & \textbf{r} \\ \hline
Solar Orbiter & (0.94, -0.24, -0.24) & 42.2$^\circ$ & 1008.5 km/s & 109.6 km/s & 5.0 & 3.2 \\ 
\hline
Wind & (0.80, -0.27, -0.53) & 22.8$^\circ$ & 853.4 km/s & 43.2 km/s & 8.2 & 3.3 \\ 
\hline
\end{tabular}
\caption{Shock Parameters derived from in-situ data at positions of Solar Orbiter and Wind in RTN coordinates, with detailed descriptions in Appendix \ref{sec:shock_parameters}. }
\label{tab:shock}
\end{table}

\subsection{Electron acceleration at the boundary layer of the IIS}

\begin{figure}[htbp]
\centering
\includegraphics[width=0.9\linewidth]{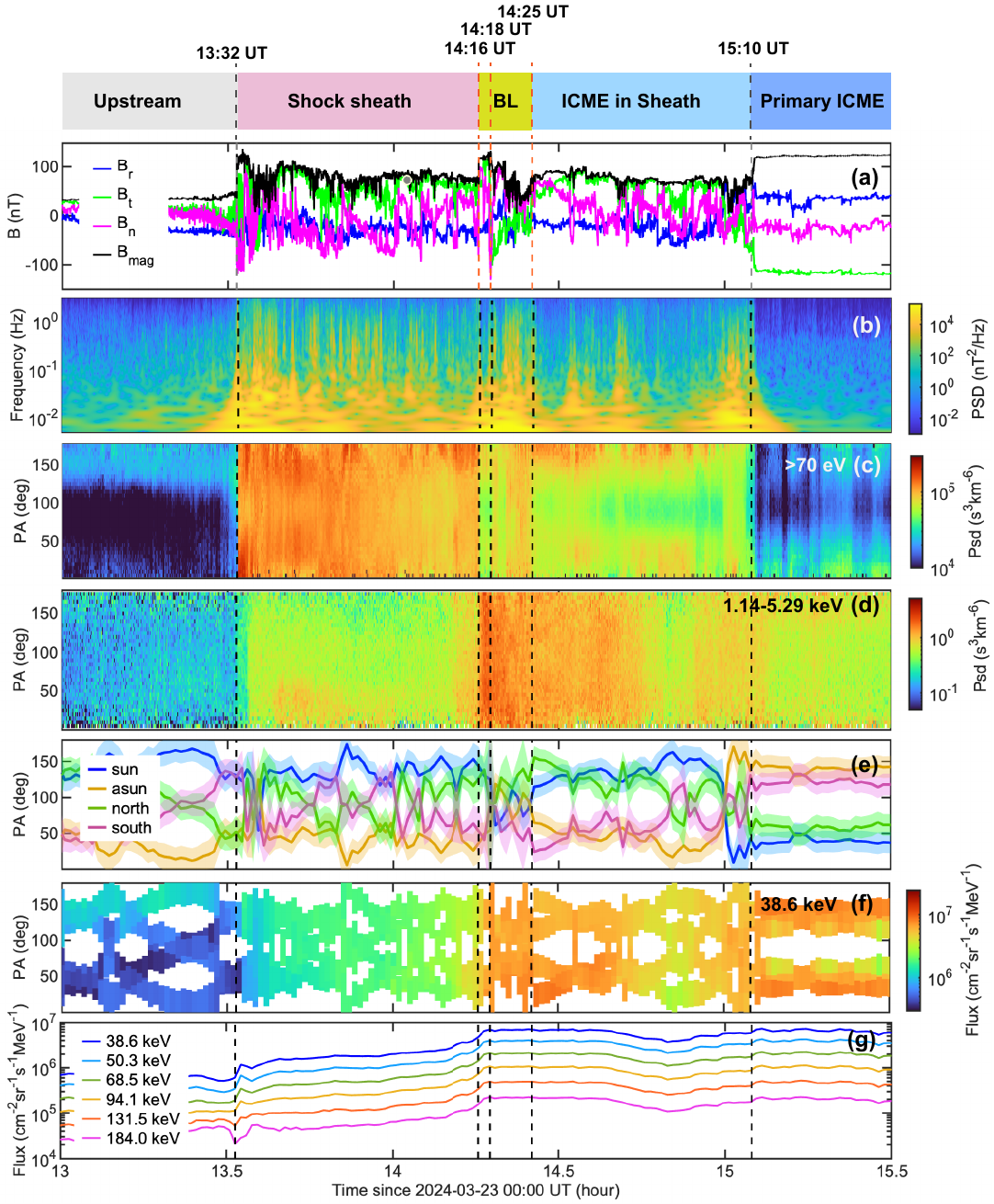}
\caption{\label{fig:electron} Magnetic fields observed by MAG and electron observation by EAS and EPT. (a) Magnetic field in RTN coordinates. (b) wavelet PSD of magnetic field fluctuations. (c) PADs of strahl electrons ($\textgreater$70 eV) (d) PADs of electrons over 1.14-5.29 keV. (e) PAs of EPT dectectors (f) PADs of electrons at 38.6 keV. (g) temporal profiles of electrons over 38-186 keV averaged in four directions. }
\end{figure}

To investigate the electron acceleration mechanisms in the IIS, we analyze electron observations from EAS and EPT. In this study, we refer to the electrons observed by SWA-EAS as strahl electrons ($\textgreater$70 eV) and suprathermal electrons ($\textgreater$1 keV), and to those observed by EPT as energetic electrons ($\textgreater$38 keV), with this classification based on the respective energy ranges of the two instruments. Panel (b) in Fig. \ref{fig:electron} shows magnetic field fluctuations in the frequency range of 0.01–3 Hz, derived from a continuous wavelet transform of the magnetic field. Enhanced large-scale fluctuations are observed at the shock front, the IIS boundary, and the boundary between the IIS and the primary ICME. Alfv$\acute{e}$nic fluctuations at higher frequencies ($\textless$3 Hz) are stronger in the shock sheath than in the IIS. In Fig. \ref{fig:electron}, Panels (c) and (d) show the PADs of strahl electrons ($\textgreater$70 eV) and suprathermal electrons (1.14-5.29 keV), respectively. In the normal shock sheath, strahl electrons exhibit enhanced flux and are nearly isotropic, while in the IIS they show bi-directional streaming with reduced intensity. Suprathermal electrons ($\textgreater$1 keV) remain nearly isotropic in both the shock sheath and the IIS, with a slight increase near the shock front followed by a pronounced enhancement at the IIS boundary and a gradual decrease within the IIS. Panel (e) shows the PAs of four EPT detectors, while Panel (f) presents the PADs of 38.6 keV electrons observed by EPT. In the shock upstream, electrons at 38.6 keV are primarily sunward-directed. Within both the shock sheath and the IIS, they were nearly isotropic. After entering the primary ICME, the electrons exhibit bi-directional streaming along sunward and anti-sunward directions. Panel (g) displays the temporal profiles of electrons over 38–186 keV, averaged over four directions. The flux increases slightly at the shock front, then gradually rises in the shock sheath, followed by a sharp enhancement at the IIS boundary around 14:16 UT, forming a $\sim$30-minute plateau. Afterwards, the flux increases again and remain high within the primary ICME, along the sunward and anti-sunward directions. Appendix \ref{sec:ept} provides the corresponding full-day plot for 2024 March 23 of Fig. \ref{fig:electron}, where bi-directional electrons at 38.6 keV inside the primary ICME can be seen more clearly. As illustrated in Panel (c) of Fig. \ref{fig:sw_mag}, Solar Orbiter likely crossed one leg of the primary ICME, suggesting that the bi-directional electrons were mirrored along the MFR.

\begin{figure}[htbp]
\centering
\includegraphics[width=0.9\linewidth]{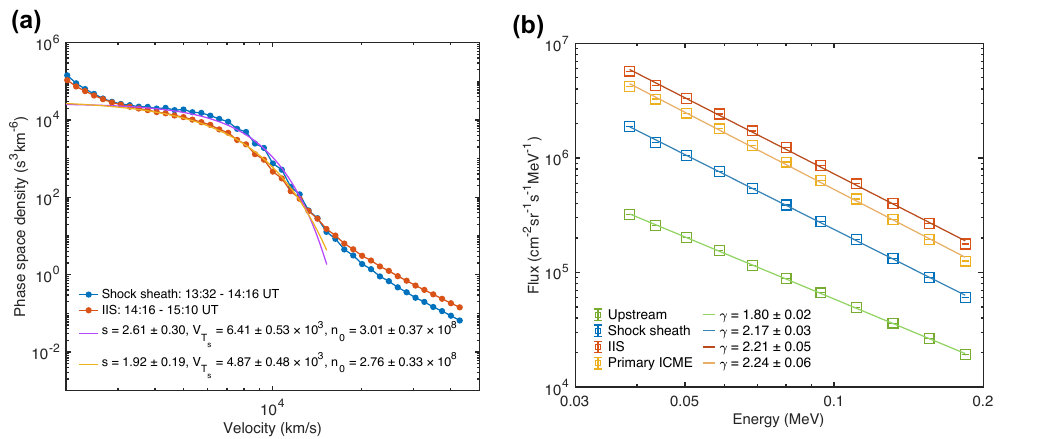}
\caption{\label{fig:electron_spectrum} Panel (a) shows the electron VDFs observed by EAS. Panel (b) shows the electron spectra observed by EPT. }
\end{figure}

As shown in Panel (a) of Fig. \ref{fig:electron_spectrum}, we then analyze the velocity distribution functions (VDFs) of solar wind electrons within the shock sheath and the IIS. The VDFs are averaged over all directions. The VDFs in both the shock sheath and the IIS presented typical flat-top distribution, which is fitted with the one-dimensional self-similar distribution function of \cite{vdf}: $f_s(v,t)=\frac{n_0s}{2V_{T_s}\Gamma(\frac{1}{s})}e^{{-\frac{v}{V_{T_s}}}^s}$, where $n_0$ (in $km^{-3}$) is the number density,  $V_{T_s}$ (in km/s) is the thermal speed and s determines the exponential shape of the distribution. When s $\xrightarrow{}$2, the distribution returns to the Maxwellian distribution. The fitted results are presented in Panel (a). Although our analysis focuses on electrons above 20 eV, where the influence of spacecraft potential is reduced, it can still slightly alter the derived VDFs and introduce minor uncertainties in interpretation. Compare to the IIS, The VDFs in the shock sheath have larger exponents, indicating flatter shapes over $\sim$20-200 eV and steeper slopes at higher energies. These coincide with intense low-frequency Alfv$\acute{e}$n waves ($\textless$3 Hz) in the shock sheath, suggesting possible wave–particle interactions \citep{flat_top}. Higher-frequency fluctuations are not further analyzed in this study. 

Electrons above 1 keV exhibit a power-law tail, with higher fluxes in the IIS than in the sheath. As shown in Panel (d) of Fig. \ref{fig:electron}, the electron flux above 1 keV peaks at the IIS boundary and then gradually decreases inside the IIS. These features suggest possible local acceleration at the IIS boundary. Panel (b) of Fig. \ref{fig:electron} shows enhanced magnetic field fluctuations within the IIS boundary, which may scatter electrons efficiently and explain the isotropy of electrons above 1 keV. Assuming electrons are accelerated by magnetic reconnection at the IIS boundary, which is similar to electron acceleration observed at the ICME boundary in \cite{prl}, the sharp magnetic field variation at 14:18 UT may generate multiple small-scale current sheets and magnetic islands, further accelerating electrons via electric field and second-order Fermi processes.

We next analyze the energetic electron spectra observed by EPT over 38–186 keV. While solar wind electrons at $\sim$1 keV are likely part of the continuous outflow of coronal electrons, electrons above 38 keV are generally associated with solar eruptions. As shown in Panel (g) of Fig. \ref{fig:electron}, plateau-like flux enhancements are observed within both the IIS and the primary ICME, likely associated with flare-accelerated electrons injected from the footpoints of the two MFRs. If the electrons originate from solar eruptions occurred during the Solar Orbiter passage through the IIS and the primary ICME, they would exhibit velocity dispersion, which is often used as a diagnostic of magnetic cloud topology \citep{electrons_mc_vda1,electrons_mc1}. However, no other eruptions were recorded at AR3614 on 2024 March 23, and no velocity dispersion is observed during passage of the IIS and primary ICME. It suggests that the plateau-like enhancements in the IIS and the primary ICME likely originate from flare-related acceleration during the eruptions of the two MFRs. For the IIS, the electron population was likely enhanced by additional injections as the primary ICME-driven shock propagated through the IIS, along with local acceleration at the IIS boundary. 

To further explore the energetic electron population in the IIS, we derived and compared electron spectra for four periods: the shock upstream from the MFR eruption at 01:00 UT until the shock arrival at 13:32 UT; the shock sheath from 13:32 UT to 14:16 UT; the IIS from 14:16 UT to 15:10 UT; and the primary ICME from 15:10 UT to 20:15 UT. Panel (b) of Fig. \ref{fig:electron_spectrum} shows energetic electron spectra observed by EPT after background subtracted, with background defined as the mean flux prior to the solar eruption. The spectra are fitted with a single power-law function. Notably, the spectral indices of the shock sheath, the IIS and the primary ICME are all close to $\sim$2.2. The spectral index in the shock upstream is slightly smaller, likely due to the transport effect. The similar spectral indices inside the shock sheath, the IIS and the primary ICME suggests that electron population in these regions has a similar origin. This indicates that the energetic electrons in the IIS also mainly come from acceleration processes near the sun, while the contribution of the shock-ICME interaction and local acceleration in the IIS is minor and likely masked by the high background flux.      

\subsection{Protons and heavy ions in the shock sheath}

\begin{figure}[htbp]
\centering
\includegraphics[width=1\linewidth]{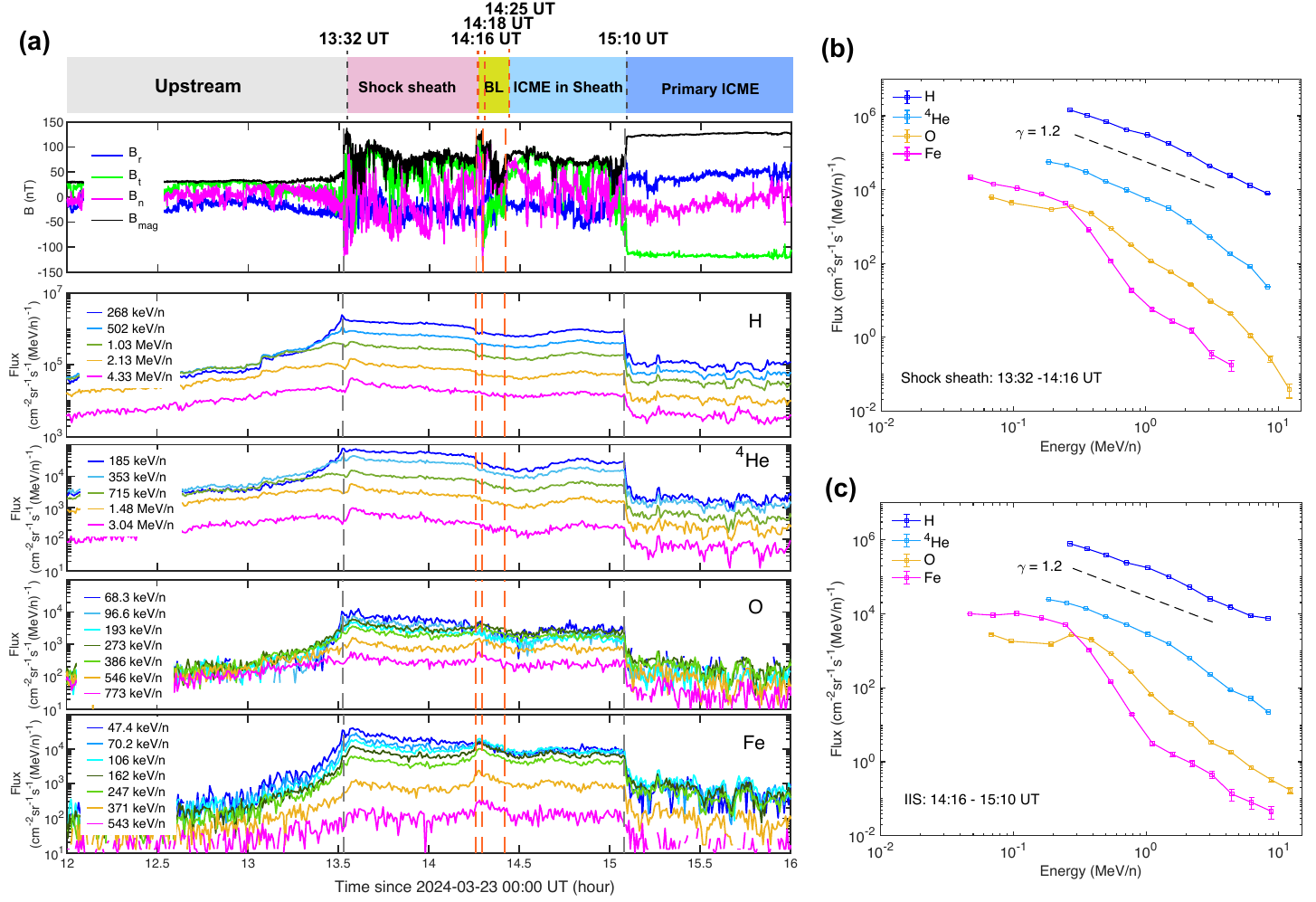}
\caption{\label{fig:element_profile} Panel (a) presents temporal profiles of H, $^4$He, Fe and O observed by SIS averaged over a and b directions. Panel (b) shows the particle spectra in the shock sheath. Panel (c) shows the particle spectra in the IIS.}
\end{figure}

Panel (a) of Fig. \ref{fig:element_profile} shows the time–intensity profiles of protons and heavy ions observed by SIS, averaged over directions a (sunward) and b (anti-sunward). At the time of shock arrival, low-energy particle fluxes (e.g., protons at 268 keV/n) peaked immediately, while higher-energy fluxes peaked about 3 minutes later. At the IIS boundary, the fluxes of protons and $^4$He dropped sharply, whereas those of Fe and O exhibited a slight enhancement, suggesting a minor contribution from the local acceleration at the boundary. Overall, the fluxes of protons and heavy ions in the IIS are lower than in the shock sheath, but significantly higher than in the primary ICME. It suggests that the fluxes in the IIS are probably contributed by the interaction processes when the primary ICME-driven shock propagated through the IIS. In this scenario, shock-accelerated particles could diffuse within the IIS, while the shock also further accelerated particles confined in the IIS.        

The spectra of H, $^4$He, Fe, and O in the sheath and the IIS are shown in Panels (b) and (c) of Fig. \ref{fig:element_profile}, respectively. The background was determined from the mean flux prior to the solar eruption and has been subtracted. According to the diffusive shock acceleration (DSA; \citealt{dsa}) theory, the spectral index $\gamma$ is related to the density compression ratio $r$ by $\gamma = \frac{r+2}{2(r-1)}$. With a compression ratio of approximately 3.2, the predicted spectral index is about 1.2. In both the sheath and the IIS, the spectra of protons and $^4$He are generally consistent with the DSA prediction, indicating that the compression ratio probably did not undergo significant variation as the shock propagated through the IIS and merged with the preceding ICME-driven shock.   

In contrast, the Fe and O spectra exhibit a double power-law shape, with a break energy around 200–300 keV/n. Below the break energy, Fe and O display flat spectra, especially within the IIS. As shown in Panel (a) of Fig. \ref{fig:element_profile}, the temporal profiles of Fe and O at low energies (~50–200 keV/n) exhibit overlapping curves within the IIS. Below 200 keV/n, the Fe/O ratio is $\sim$2 in the sheath but rises to $\sim$8 in the IIS, indicating a contribution of superthermal particles originating from the solar eruption. In both the normal shock sheath and the IIS, above the break energy, the O spectrum is slightly softer than predicted by DSA, while the Fe spectrum is significantly softer. Correspondingly, the Fe/O ratio drops sharply with energy, reaching $\sim$0.05 above 700 keV/n in both regions. This behavior likely reflects the less efficient acceleration of high-rigidity ions \citep{Fe_O_spectrum}, as Fe is accelerated less efficiently than O at higher energies, resulting in a much softer spectrum.

\section{Conclusion and discussion} \label{sec:summary}

In this study, we present a comprehensive analysis of an IIS structure formed during a twin-CME event on 2024 March 23. Solar Orbiter (at 0.4 AU) and Wind (at L1 point) were radially aligned during the event and both observed the IIS structure. As the shock propagates from 0.4 AU to 1 AU, its speed decreases while the compression ratio remains nearly unchanged and the Alfv$\acute{e}$n Mach number increases, transitioning from an oblique shock at Solar Orbiter to a quasi-parallel shock at Wind. At 0.4 AU, the IIS observed by Solar Orbiter showed clear bi-directional electron streaming and a stable magnetic field, lasting for approximately 54 minutes. A stream interface is identified at the IIS boundary, characterized by sharp magnetic field variations and strong flow shear. At 1 AU, the IIS observed by Wind exhibited a typical MFR-like magnetic structure with a duration of about 2.4 hours. Compared with the IIS boundary at 0.4 AU, the IIS boundary at 1 AU appeared thinner and more moderate. During its interplanetary propagation, the IIS underwent significant expansion. Meanwhile, the dominant magnetic component of the IIS shifted from $B_t$ at 0.4 AU to $B_n$ at 1 AU, suggesting a possible MFR rotation. It is also worth noting that, due to the deflection of the IIS during its propagation, Solar Orbiter and Wind may have crossed different parts of the structure, making a direct comparison difficult.

We propose two possible scenarios to explain the differences in the magnetic field configuration of the IIS observed by Solar Orbiter and Wind. The first involves shock compression of the MFR, which can modify its magnetic field and lead to the formation of magnetic discontinuities that account for the thick and complex boundary structure observed at 0.4 AU. This compression may also trigger magnetic reconnection, further distorting the MFR configuration. We suggest that the IIS at 0.4 AU was observed shortly after being overtaken by the shock. Numerical simulations of shock–ICME interactions \citep{shock_icme_xiong} also show magnetic field compression and rotation during the interaction, supporting this interpretation. The second scenario considers interactions between the IIS and the surrounding structures, including the shock sheath and the primary ICME, during its propagation. These processes may involve moderate reconnection with the shock sheath and compression or squeezing by the primary ICME, resulting in minor changes to the IIS structure.

We analyzed the pitch-angle distributions and energy spectra of electrons ranging from $\sim$70 eV to 186 keV within both the shock sheath and the IIS structure. In the shock sheath, strong turbulence produced nearly isotropic electron populations at these strahl energies with flat-top VDFs, whereas in the IIS, the strahl eletrons showed bi-directional streaming and less distinct flat-top features. Suprathermal electrons ($\textgreater$1 keV) were isotropic in both regions, with a clear flux enhancement at the IIS boundary followed by a gradual decrease inside the IIS, indicating possible local acceleration at the boundary. Such acceleration may result from magnetic reconnection at the boundary, with interactions with magnetic islands serving as an effective acceleration mechanism \citep{reconnection_Drake,reconnection_zank}. For energetic electrons (38–186 keV), we observed a sustained flux enhancement within the IIS. The electron spectrum within the IIS follows a single power law with a spectral index of $\sim$2.2, consistent with the spectral indices in the shock sheath and the primary ICME, suggesting a common solar origin. Although local acceleration at the IIS boundary may contribute, its effects appear to be masked by the dominant flare-related population. \cite{kh} discussed how the Kelvin–Helmholtz (K–H) instability at the CME boundary during the intense geomagnetic storm on 2024 May 10 could distort the magnetic field and trigger local small-scale reconnection events. This storm was associated with multiple interacting CMEs \citep{Storm_May_Liu}, suggesting that such local acceleration processes may commonly occur at the boundaries of strong CMEs, especially during CME–CME interactions. Future statistical studies on IIS structures could further verify this. 

The energetic ion observations suggest that the fluxes within the IIS are primarily contributed by interaction processes occurring as the primary ICME-driven shock propagated through the preceding ICME. The spectra of protons and $^4$He in both the shock sheath and the IIS are generally consistent with predictions from DSA, indicating that the shock strength likely remained stable as it traversed the IIS. In both the shock sheath and the IIS, Fe and O exhibit a double power-law spectra with a break at $\sim$200–300 keV/n. Below the break, their spectra are very flat, particularly in the IIS. It suggests the presence of seed particles that have not been fully accelerated into a power-law distribution. This may be attributed to the limited acceleration time as the shock rapidly traversed the IIS, as well as a possible reduction in shock strength within the IIS \citep{Lugaz_simulation}, both of which could lead to decreased acceleration efficiency. The elevated Fe/O ratio ($\sim$8) at lower energies in the IIS suggests a contribution from flare-related suprathermal particles. Compared to DSA predictions, the Fe and O spectra are significantly softer, especially for Fe, and the sharp Fe/O drop at high energies suggests less efficient acceleration of high-rigidity ions.  


\begin{acknowledgements}

We thank the anonymous referee for the valuable comments and suggestions, which greatly improved the manuscript. We acknowledge the use of data from Solar Orbiter, Wind, CHASE, SDO and SOHO spacecrafts. This work is supported  by NSFC under grant 12333009 and by the Fundamental Research Funds for the Central Universities under grant KG202506. Solar Orbiter is a mission of international cooperation between ESA and NASA, operated by ESA. Solar Orbiter SWA work at UCL/MSSL is currently funded under UKRI and UKSA grants UKRI1204, ST/W001004/1 and UKRI1919.  

\end{acknowledgements}

\bibliographystyle{aasjournal}
\bibliography{ref}

\appendix
\renewcommand{\thefigure}{A\arabic{figure}}
\setcounter{figure}{0}

\section{Calculation of Shock Parameters} \label{sec:shock_parameters}

The method for the calculation of shock parameters presented in Table \ref{tab:shock} is referred to the methods described in \cite{shock_parameter_kilpua} and \cite{shock_parameter_walker}. We first determine the time period of the shock and treat the shock as a magnetic discontinuity. The magnetic field and plasma parameters in the upstream region denoted by the subscript ``u" are averaged over a 10-minute interval before the shock arrival. Parameters with the superscript ``up" are also averaged over this upstream interval. Similarly, the downstream parameters denoted by the subscript ``d" are averaged over a 10-minute interval after the shock arrival.

The shock normal is determined by the MD3 method \citep{md3} using both the magnetic field and solar wind data.

\begin{equation}
    n_{shock}=\frac{(B_d-B_u)\times((B_d-B_u)\times(V_d-V_u))}{|(B_d-B_u)\times((B_d-B_u)\times(V_d-V_u))|}
\end{equation}

$\theta_{Bn}$ is the separation angle between the shock normal and the upstream magnetic field. 

\begin{equation}
    \theta_{Bn}=arccos(\frac{|B_u\cdot n_{shock}|}{\|B_u\|\|n_{shock}\|})
\end{equation}

The shock velocity is calculated in the spacecraft coordinate using the mass flux conversion across the shock:

\begin{equation}
    V_{shock}=\big|\frac{n_dv_d-n_uv_u}{n_d-n_u}\cdot n_{shock}\big|
\end{equation}

The upstream Alfv$\acute{e}$n velocity is:

\begin{equation}
    V_A^{up}=\left\langle \frac{B_u}{\mu_0N_pm_p}\right\rangle _{up}
\end{equation}

The Alfv$\acute{e}$n Mach number is calculated in the rest frame of the shock:

\begin{equation}
    M_A=\frac{V_u\cdot n_{shock} \pm V_{shock}}{V_A^{up}}
\end{equation}

The density compression ratio is:

\begin{equation}
    r=\frac{N_d}{N_u}
\end{equation}

\section{Electron observations during 2024 March 23} \label{sec:ept}

\begin{figure}[htbp]
\centering
\includegraphics[width=0.9\linewidth]{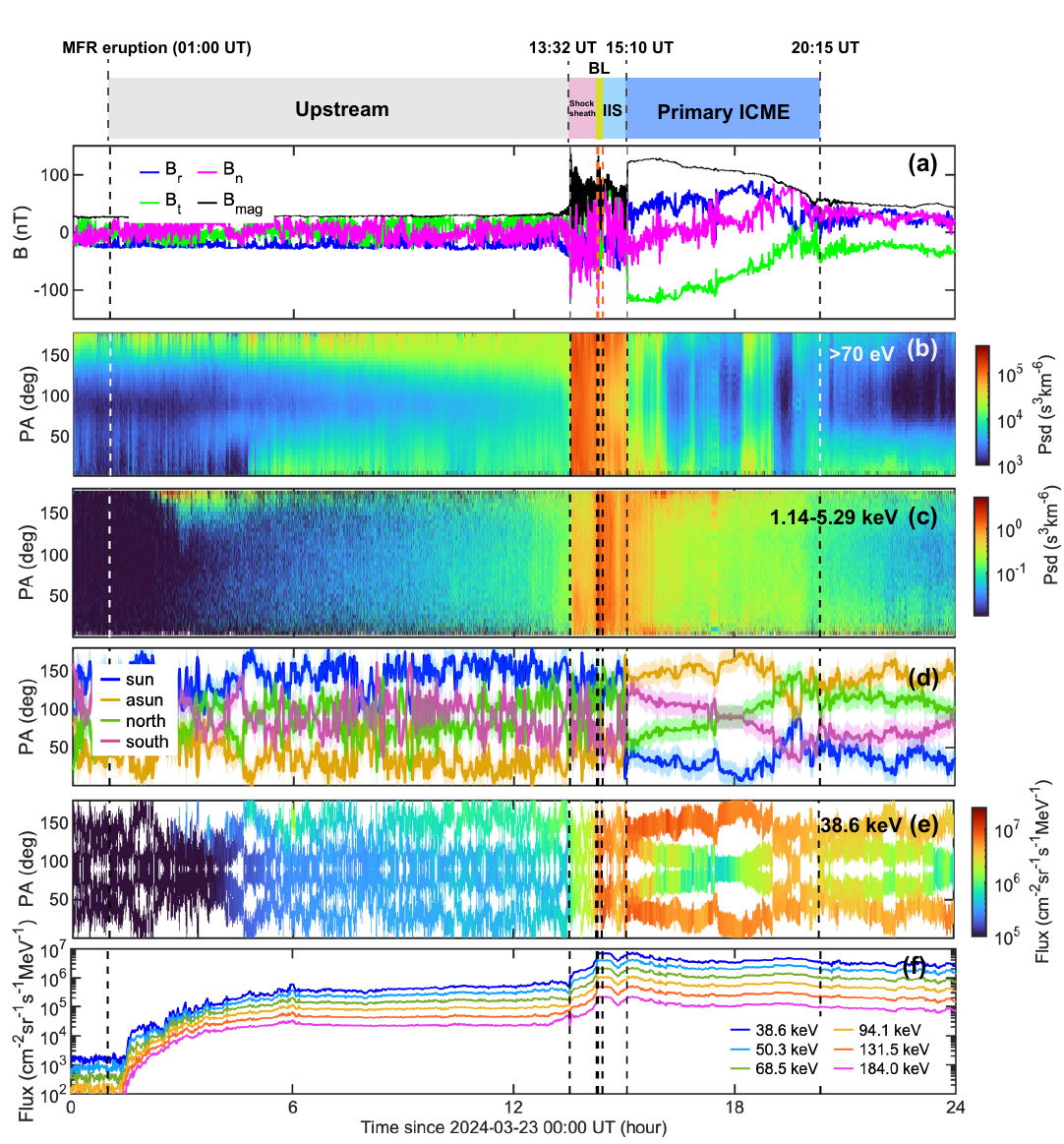}
\caption{\label{fig:electron_whole_day} Electron observation by EAS and EPT over 2024 March 23. (a) Magnetic field in RTN coordinates. (b) PADs of strahl electrons ($\textgreater$70 eV) (c) PADs of electrons over 1.14-5.29 keV. (d) PAs of EPT dectectors (e) PADs of electrons at 38.6 keV. (f) temporal profiles of electrons over 38-186 keV averaged in four directions. }
\end{figure} 

\end{CJK}
\end{document}